\documentclass[amsmath,twocolumn,superscriptaddress,prl,aps]{revtex4}
\usepackage{amsthm,amsfonts,graphicx,verbatim}
\usepackage{graphicx}
\usepackage{bm}
\newcommand{\be}{\begin{equation}}
\newcommand{\ee}{\end{equation}}
\newcommand{\bea}{\begin{eqnarray}}
\newcommand{\eea}{\end{eqnarray}}

\newcommand{\Tr}{{\rm Tr}\,}
\renewcommand{\epsilon}{\varepsilon}
\renewcommand{\vec}[1]{{\bf #1}}

\renewcommand{\cite}[1]{[\onlinecite{#1}]}

\newcommand{\sign}{\mathrm{sign}}

\begin{document}

\title{Interplay of superconductivity and spin density wave order in doped graphene}
 \author{Rahul Nandkishore}
 \affiliation{Department of Physics, Massachusetts Institute of Technology, Cambridge MA 02139, USA}
\author{Andrey V. Chubukov}
\affiliation{Department of Physics, University of Wisconsin-Madison, Madison, WI 53706, USA}

\begin{abstract}
We study the interplay between superconductivity and spin density wave order in graphene doped to
 $3/8$ or $5/8$ filling (a Van Hove doping). At this doping level, the system is known to exhibit weak coupling instabilities to both chiral $d+id$ superconductivity and to a uniaxial spin density wave.
 Right at van Hove doping, the superconducting instability is strongest and emerges at the highest $T_c$, but slightly away from van-Hove doping a spin-density-wave likely emerges first.   We investigate whether
at some lower temperature superconductivity and spin-density-waves co-exist.  We derive the Landau-Ginzburg functional describing interplay of the two order parameters. Our calculations show
  that superconductivity and spin density wave order do not co-exist and are separated by first-order transitions, either as a function of doping or as a function of T.
\end{abstract}

\maketitle
Two dimensional electron systems provide an ideal environment for exploration of many body physics. Graphene, as a new two dimensional electron system, may allow us to access new many body phases that have not been hitherto observed. Unfortunately, undoped single layer graphene seems to be well described by a non-interacting model \cite{CastroNeto}, with the vanishing density of states suppressing interaction effects. In order to access many body physics in graphene, one must sidestep the vanishing density of states. One way to do this is by doping.
  When graphene is doped to the $M$ point of the Brillouin zone - a doping level that corresponds to
 $3/8$ (or $5/8$) filling (undoped graphene  corresponds to $1/2$ filling)
  - the Fermi surface undergoes a topological transition from a two piece to a one piece Fermi surface \cite{Wallace}. Associated with this topological transition is a divergent density of states, which gives rise to weak coupling instabilities to unusual many body states. Doped graphene thus provides a promising playground for exploration of new quantum many body states.

 The recent success of experimental efforts to dope graphene to the $M$ point \cite{McChesney} has inspired a flurry of theoretical works studying many body physics in doped graphene \cite{Nandkishore, TaoLi, MoraisSmith, FaWang, Thomale, SDW, Z4, Gonzalez, TaoLioriginal, Vozmediano, ChiralPG}. It has been established \cite{Nandkishore} that the principal weak coupling instabilities are to chiral $d+id$ superconductivity
 and to a uniaxial spin density wave
 (SDW),
 with the superconducting
 instability leading at Van Hove doping \cite{Nandkishore}, and the
 SDW
 leading somewhat away from Van Hove doping \cite{SDW}. It is not known, however, whether these two orders are mutually exclusive, or whether they co-exist in some range of temperatures and dopings.

In this letter we demonstrate that for graphene near the $M$ point, superconductivity and
 SDW
magnetism are  mutually exclusive orders.
 We derive Landau-Ginzburg action for two order parameters and show that the interplay between quartic terms is such that the minimum of the
 action  is when only one order
  parameter is  nonzero. This result
  stands in stark contrast to pnictide materials, where
  Landau-Ginzburg analysis shows that
   superconducting and SDW
     orders  do
  co-exist \cite{Vorontsov, Schmalian}. In doped graphene one expects to observe pure chiral superconductivity at Van Hove doping, with a
   first-order
   transition to pure spin density wave order upon doping away from the Van Hove point. Our conclusions apply also to doped triangular lattice systems \cite{Martin+Batista}, which have an identical low energy description
    near $3/4$ filling.

\begin{figure}
\includegraphics[width = 0.6\columnwidth]{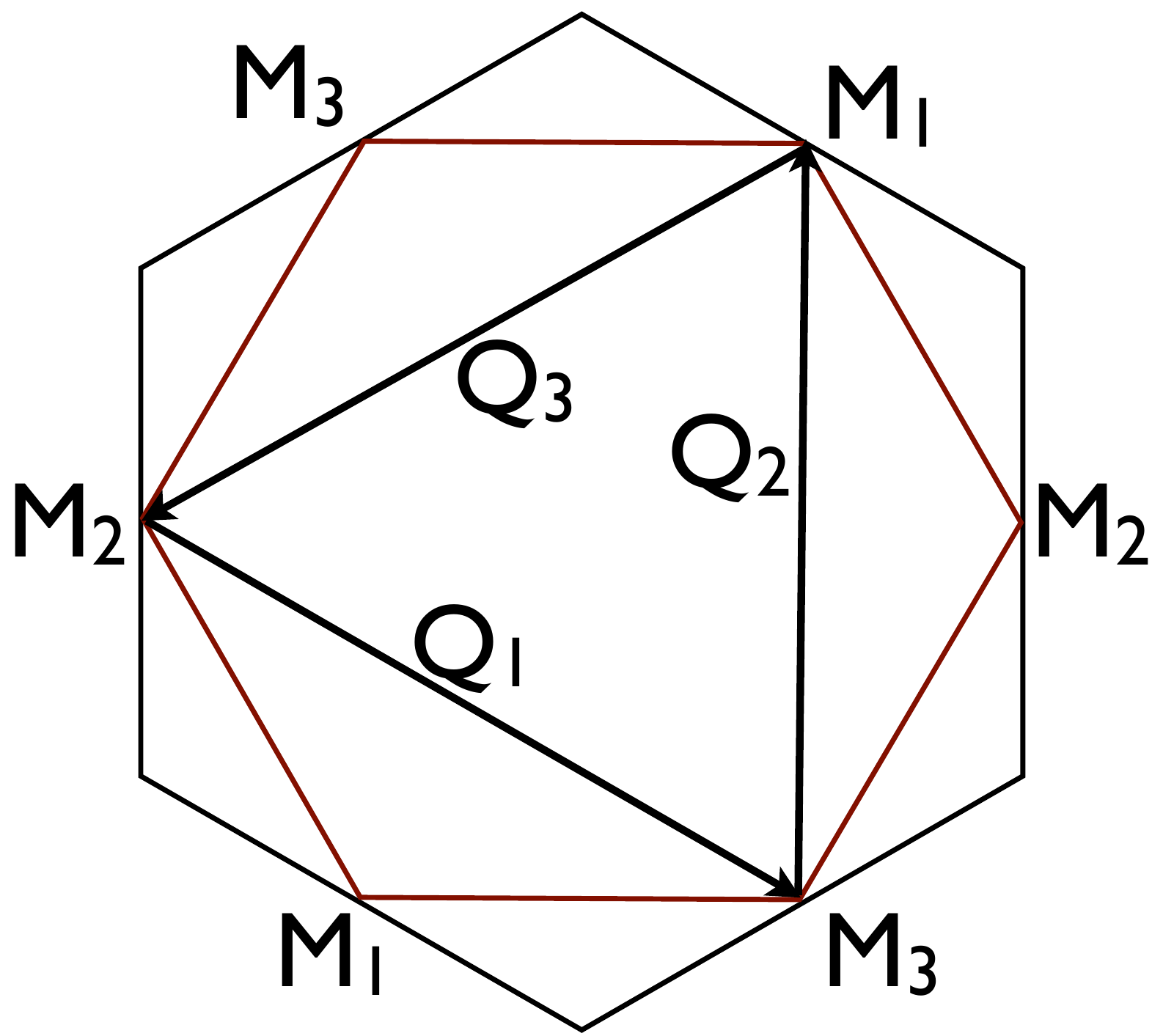}
\caption{\label{fig: FS} The Fermi surface at Van Hove doping is a perfect hexagon inscribed within the hexagonal Brillouin zone. The hexagon has three inequivalent corners, labeled $M_{1,2,3}$, which are saddle points of the dispersion
 and give rise to a divergent density of states. Each saddle point is perfectly nested with each other saddle point. The perfect nesting of the FS
  is broken only by third and higher neighbor hoppings, which are generally quite small. Meanwhile, the existence of saddle points is fully robust, being a consequence of a topological transition from a 
  Fermi surface with two inequivalent pieces 
 to a one piece Fermi surface}
\end{figure}

{\it The model:}  Our point of departure is the tight binding model \cite{Wallace}, with the nearest-neighbor dispersion
\begin{equation}
\label{eq: dispersion}
\epsilon_{\vec{k}} = \pm t \sqrt{1 + 4 \cos{\frac{k_y \sqrt{3}}{2}} \cos{\frac{3 k_x}{2}} + 4 \cos^2{\frac{k_y \sqrt{3}}{2}}} - \mu
\end{equation}
 where the overall sign is $+$ or $-$ depending on whether we are above or below half filling.
 For definiteness we take a plus sign.
   Van Hove doping 
    then 
    corresponds to
   $\mu = t$,
    at which point the Fermi surface has the form shown in (Fig. \ref{fig: FS}).
 The Fermi velocity vanishes near the hexagon corners ${\bf M_1} = (2\pi/3,0),~{\bf M_2} = (\pi/3,\pi/\sqrt{3}),~{\bf M_3} = (-\pi/3,\pi/\sqrt{3})$,  which are saddle points of the dispersion:
\begin{eqnarray}
\epsilon_{k \approx M_1} &=& \frac{3t}{4}(3k_x^2 - k_y^2), \qquad
\epsilon_{k \approx M_2} = \frac{3t}{4} 2k_y(k_y - \sqrt{3}k_x), \nonumber\\
\epsilon_{k \approx M_3} &=& \frac{3t}{4} 2k_y(k_y + \sqrt{3}k_x)
\label{n_1}
\end{eqnarray}
Each time $k$ is a deviation from a saddle point.
 Saddle points give rise to a logarithmic singularity in the DOS and control physics at weak coupling.
  There are three in-equivalent nesting vectors $Q_{a b}$  connecting
in-equivalent pairs of saddle points ${\bf M}_a$ and ${\bf M}_b$ (see Fig.\ref{fig: FS}):
 \begin{eqnarray}
 Q_1 &=&  Q_{23} = (\pi, \pi/\sqrt{3}); \quad Q_2 = Q_{31} = (\pi, -\pi/\sqrt{3}) \nonumber\\
 Q_3 &=& Q_{12} = (0, 2\pi/\sqrt{3}).
\end{eqnarray}
  Each $\vec{Q}_{i}$ is physically the same as $- \vec{Q}_{i}$ because $\vec{Q}_{i}$ is half of a reciprocal lattice vector.

There are four different interactions between fermions near saddle points, $g_i$, $i=1-4$, with momentum transfer near zero and near $Q_i$.
(Refs. \onlinecite{Nandkishore, FaWang, Thomale, SDW, Z4, Gonzalez, Vozmediano, ChiralPG}).
For our purposes, relevant interactions are density-density interaction within one patch ($g_4$) and between patches ($g_2)$ and
 the interaction which describes hopping of a pair of fermions from one patch to the other ($g_3$).  The fourth interaction $g_1$
  is the exchange interaction between patches.
 Interactions $g_2$ and $g_3$ renormalize particle-hole vertices and control the SDW instability, while interactions $g_3$ and $g_4$
 renormalize particle-particle vertices and control the superconducting instability (note that $g_3$ contributes to both instabilities).

  The partition function $Z$
  of the model can be written
  as a functional integral over Grassman valued (fermionic) fields $\psi$. We have $Z = \int D[\bar \psi, \psi] \exp\big(-S(\bar \psi, \psi) \big)$, where $S = \int_0^{1/T} \mathcal{L}(\vec{k},\tau)$ ($T$ is the temperature) and
\begin{widetext}
\begin{equation}
\mathcal{L} = \sum_{a \alpha} \bigg[\bar \psi_{a,\alpha} \big(\partial_{\tau} + \epsilon_{\vec{k}} - \mu - g_4 \bar \psi_{a,\bar \alpha}
 \psi_{a,\bar \alpha}\big) \psi_{a,\alpha} - \sum_{b \neq a} \sum_{\beta}  g_1  \bar \psi_{a,\alpha}\bar \psi_{b,\beta} \psi_{a,\beta} \psi_{b, \alpha}  +
  g_2 \bar \psi_{a,\alpha}\bar \psi_{b,\beta}
 \psi_{b,\beta} \psi_{a, \alpha} + g_3 \bar \psi_{a,\alpha}\bar \psi_{a,\beta}  \psi_{b,\beta} \psi_{b,\alpha}
 \bigg]. \label{eq: action}
\end{equation}
\end{widetext}
 Here 
  $a, b = 1,2,3$ label which saddle point we are closest to, $\alpha$ and $\beta$ are spin labels, and $\bar \alpha$ is the opposite spin state to $\alpha$. We have retained only those states that are close to the saddle points
 - this `patch model' is exact in the limit of weak coupling \cite{Nandkishore}.

This action displays instabilities towards d-wave
 superconductivity and SDW.
  We therefore decouple the interactions in the d-wave superconducting and
   SDW channels simultaneously, by means of two Hubbard Stratanovich transformations. We introduce the Hubbard-Stratanovich
   superconducting
   fields $\Delta_a = (g_3 - g_4) \langle \psi_{a, \uparrow} \psi_{a, \downarrow}\rangle$. Since the superconductivity is known to be $d+id$ \cite{Nandkishore}, we
    set  $(\Delta_1, \Delta_2, \Delta_3) = \Delta (1, e^{2i\pi/3}, e^{-2i\pi/3})$ and describe superconducting
   fields by a single  complex order parameter $\Delta$.
    We also introduce the
    three SDW
    order parameters $M_{ab} = (g_2+g_3) \langle \bar \psi_{a, \uparrow} \psi_{b, \downarrow} \rangle$.
    Since the
     SDW order
     is known to be uniaxial \cite{SDW}, we can replace the three vector order parameters $M_{12}, M_{23}, M_{31}$ by a single scalar SDW order parameter $M$, which represents the magnetic order along the SDW axis. Since the system has $O(3)$ spin rotation symmetry, the SDW axis can be chosen to coincide with the $z$ axis without loss of generality. Finally, we introduce the Nambu spinor $\chi_a$, a four component spinor defined according to $\chi_a = (\psi_{a, \uparrow}, \psi_{a, \downarrow}, \bar \psi_{a, \downarrow}, - \bar \psi_{a, \uparrow})$. The action after Hubbard Stratanovich transformation can be written in the Nambu spinor basis as
\begin{eqnarray}
\mathcal{L} &=& \frac{M^2}{g_2 + g_3} + \frac{|\Delta|^2}{g_3-g_4} + \sum_{a b} \bar \chi_{a} \mathcal{G}^{-1}_{ab} \chi_b \\
\mathcal{G}^{-1}_{aa} &=& \big(\partial_\tau 1_2 + (\epsilon_\vec{k} - \mu) \sigma_3  + \Delta \sigma_+  + \Delta^* \sigma_- \big)\otimes 1_2 \nonumber\\
\mathcal{G}^{-1}_{a\neq b} &=& M (1_2 \otimes \eta_3) \label{eq: fullgf}
\end{eqnarray}
Here the $\sigma_i$ are Pauli matrices acting in the particle-hole space, the $\eta_i$ are Pauli matrices acting in the spin space, $1_2$ is a two dimensional identity matrix, and $\sigma_{\pm} = \sigma_1 \pm \sigma_2$. The notation we have used is borrowed from \cite{ReadGreen}.

We can now integrate out the fermions exactly
 to obtain an action purely in terms of the superconducting and SDW order parameter fields,
\begin{equation}
\mathcal{L} = \frac{M^2}{g_2 + g_3} + \frac{|\Delta|^2}{g_3-g_4} - \Tr \ln \mathcal{G}^{-1}(\Delta, M) \label{eq: F}
\end{equation}
where the trace goes over Nambu spinor indices, and also over imaginary time and over momentum.
We now define $\vec{G}$ to be the
 "bare"
(matrix) Green function evaluated at $\Delta = 0, M = 0$, and
 define matrix order parameters $\vec{\Delta}$ and $\vec{M}$, such that $\mathcal{G}^{-1} = \vec{G}^{-1} + \vec{\Delta} + \vec{M}.$ We can then write
\begin{eqnarray}
\Tr \ln \mathcal{G}^{-1} &=& \Tr \ln \bigg(\vec{G}^{-1}( 1 + \vec{G} \big(\vec{\Delta} + \vec{M} \big) \bigg)\nonumber \\
&=& \rm{constant} + \Tr \ln \bigg( 1 + \vec{G} \big(\vec{\Delta} + \vec{M} \big) \bigg). \label{eq: preexpansion}
\end{eqnarray}
It is convenient to explicitly write out the expressions for $\vec{G}, \vec{\Delta}$ and $\vec{M}$. We adopt the shorthand $F^{\pm}(\omega_n, \vec{k}) = 1/\big(i\omega_n \pm (\epsilon_{\vec{k}} - \mu)\big)$, where $\omega_n = (2n+1) \pi T $ are fermionic Matsubara frequencies. Using the shorthand, we can define the various matrices as
  \begin{widetext}
\begin{eqnarray}
\vec{G} &= &\left( \begin{array}{cccccc}F^+(\omega_n,\vec{k}) & 0 & 0 & 0 & 0 & 0\\ 0& F^-(\omega_n,\vec{k}) & 0 & 0 & 0 & 0 \\ 0 & 0 & F^+(\omega_n,\vec{k+Q_1}) & 0 & 0 & 0\\ 0&0&0&F^-(\omega_n,\vec{k+Q_1})&0&0\\ 0&0&0&0&F^+(\omega_n,\vec{k+Q_2})&0\\0&0&0&0&0&F^-(\omega_n\vec{k+Q_2}) \end{array}\right)  \otimes 1_2; \nonumber\\
\vec{\Delta} &= &  \left( \begin{array}{cccccc}0 & \Delta & 0 & 0 & 0 & 0\\ \Delta^*& 0 & 0 & 0 & 0 & 0 \\ 0 & 0 &0& \Delta e^{2i\pi/3} & 0 & 0\\ 0&0&\Delta^* e^{-2i\pi/3}&0&0&0\\ 0&0&0&0&0&\Delta e^{-2i\pi/3}\\0&0&0&0&\Delta^* e^{2i\pi/3}&0 \end{array}\right) \otimes 1_2; \qquad 
\vec{M} = M \left(\begin{array}{cccccc}0 & 0 & 1 & 0 & 1 & 0\\ 0& 0 & 0 & 1 & 0 & 1 \\ 1 & 0 & 0 & 0 & 1 & 0\\ 0&1&0&0&0&1\\ 1&0&1&0&0&0\\0&1&0&1&0&0 \end{array}\right) \otimes \sigma_3.
\end{eqnarray}
Thus far, everything we have done has been exact. We now work close to $T_c$
 and perform a double expansion of (\ref{eq: preexpansion}) in small $|\Delta|$ and small $M$. We terminate the expansion at quartic order in both fields
  and drop all terms that are odd in powers of $\vec{M}$ or $\vec{\Delta}$
   as they vanish upon taking the trace.
   We then obtain
    for the $\Tr \ln$ term the expression
  \begin{equation*}
\Tr \bigg[ - \frac12 \big(\vec{G} \vec{\Delta} \vec{G} \vec{\Delta} + \vec{G} \vec{M} \vec{G} \vec{M}\big) - \frac14 \big(\vec{G} \vec{\Delta} \vec{G} \vec{\Delta} \vec{G} \vec{\Delta} \vec{G} \vec{\Delta} +\vec{G} \vec{M} \vec{G} \vec{M}\vec{G} \vec{M} \vec{G} \vec{M} + 4 \vec{G} \vec{M} \vec{G} \vec{M}\vec{\Delta} \vec{G} \vec{\Delta} + 2 \vec{G} \vec{M}\vec{G}\vec{\Delta} \vec{G} \vec{M}\vec{G}\vec{\Delta} \big)   \bigg].
\end{equation*}
We have made use of the fact that the trace of a product of matrices is invariant under a cyclic permutation of the matrices. Evaluating the traces and substituting back into (\ref{eq: F}) leads to the expression
\begin{equation}
\mathcal{L} = \alpha_1 (T-T_c) |\Delta|^2 + \alpha_2 (T-
T_N) M^2 + K_1 |\Delta|^4 + K_2 M^4 + 2 K_3 |\Delta|^2 M^2 \label{eq: free energy}
\end{equation}
where we have defined the expansion coefficients
\begin{eqnarray}
K_1 &= 3 T \sum_{\omega_n} \int \frac{d^2k}{(2\pi)^2} & \big(F^+(\omega_n, \vec{k})F^-(\omega_n, \vec{k})\big)^2  \\
K_2 &= 3 T \sum_{\omega_n} \int \frac{d^2k}{(2\pi)^2}& F^+(\omega_n,\vec{k})^2 F^+(\omega_n, \vec{k}+\vec{Q_1})^2
+ 2 F^+(\omega_n,\vec{k})^2 F^+(\omega_n, \vec{k}+\vec{Q_1})F^+(\omega_n, \vec{k}+\vec{Q_2}) + (F^+ \rightarrow F^-)\nonumber\\
K_3 &= 6 T \sum_{\omega_n} \int \frac{d^2k}{(2\pi)^2} & F^+(\omega_n, \vec{k})F^-(\omega_n, \vec{k})F^+(\omega_n, \vec{k+Q_1})F^-(\omega_n, \vec{k+Q_1}) \cos\frac{2\pi}{3} \nonumber\\&& + F^+(\omega_n, \vec{k})^2F^-(\omega_n, \vec{k})F^+(\omega_n, \vec{k+Q_1}) + (F^+ \leftrightarrow F^-) \nonumber
\label{ki}
\end{eqnarray}
\end{widetext}
  Terminating the expansion at quartic order in both order parameters is
 justified
 if the quadratic terms for superconducting and SDW order change sign at
  about
   the same critical temperature,
  $T_c \approx T_N$.
  Renormalization group analysis shows that the couplings which determine both superconducting and SDW instabilities diverge at the onset of the first instability upon lowering T~\cite{Nandkishore,FaWang,Thomale}.
 The critical temperatures, however, are determined by superconducting and SDW susceptibilities, which generally have different exponents (different anomalous dimensions).
 \begin{figure}
\includegraphics[width = 0.5\columnwidth]{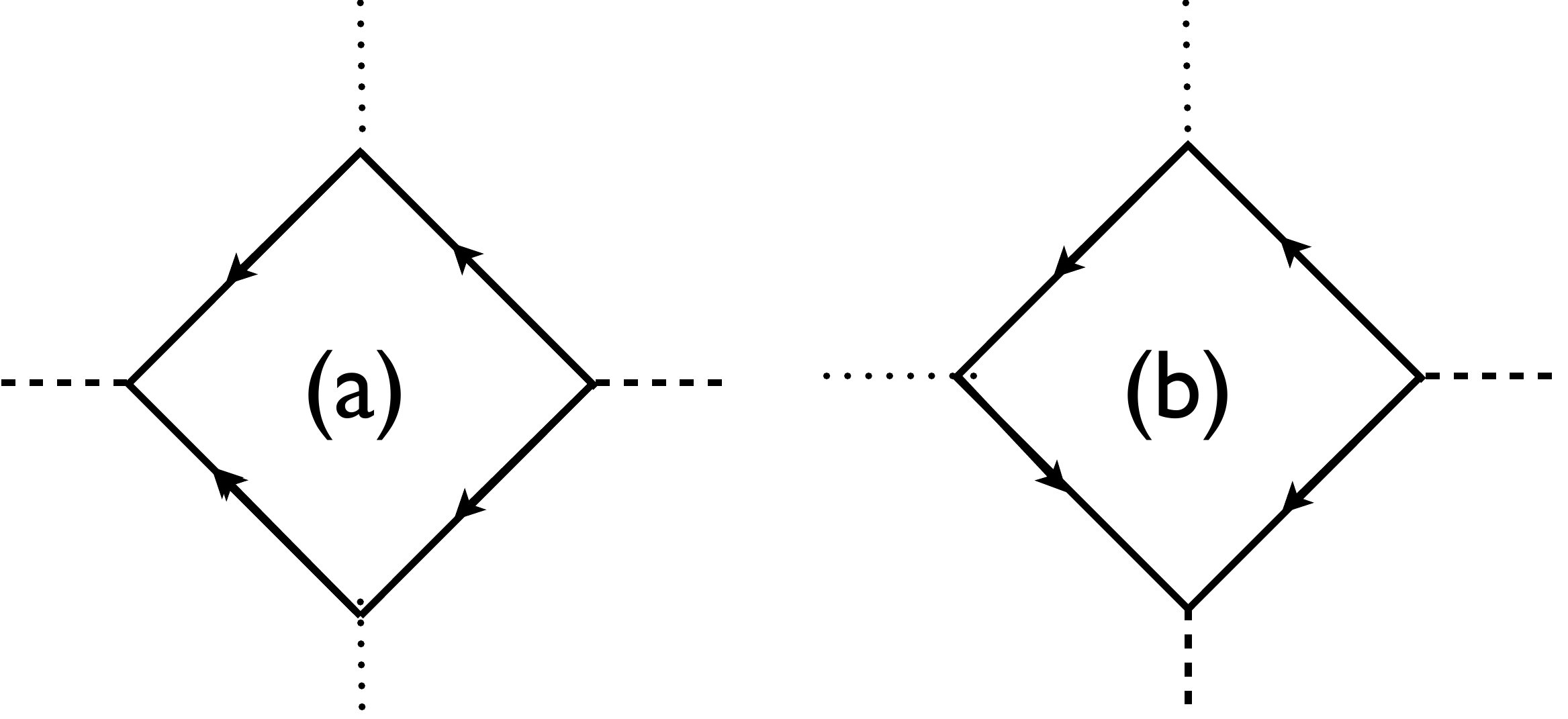}
\caption{\label{fig: diag} Diagrammatic representation of the two processes that couple superconductivity and magnetism at quartic order in
 the Landau-Ginsburg expansion.
  Lines with arrows represent fermion propagators, dotted lines represent SDW order parameter $M$, dashed lines represent superconducting order parameter $\Delta$.}
  \end{figure}
   It has been demonstrated on general grounds\cite{Vafek} that different orders emerge simultaneously when their anomalous dimensions $\eta_i >1$.
  In our case, the anomalous exponents for  superconducting and SDW susceptibilities have been calculated in~\cite{Nandkishore}.
   Using results from that work, we find that $\eta_{SC} = 1.48 >1$, while $\eta_{SDW} = 0.97$ (the results are for perfect nesting).
   Because $\eta_{SDW} <1$, $T_N < T_c$.   However, since $\eta_{SDW}$ is very close to one, we expect that $T_N$
   in (\ref{eq: free energy}) is only slightly lower than $T_c$, in which case the expansion up to quartic order in $\Delta$ and $M$ is justified.
   For dopings slightly away from the van-Hove one, we expect SDW to be the leading instability~\cite{FaWang,Thomale,SDW,Z4}. In this case, $T_N \geq T_c$, again we assume that the difference between the two critical temperatures is small.

 It was shown in the context of pnictides \cite{Vorontsov,Schmalian} that a free energy of the form (\ref{eq: free energy}) leads to co-existence
 of the two order parameters
  if $K_1 K_2 - K_3^2 >0$.
 We computed the coefficients $K_1$, $K_2$ and $K_3$ in our case by explicitly integrating over fermionic momenta and summing over fermionic frequencies in (\ref{ki}) (for details see Supplementary material). We obtain, with logarithmic accuracy
 \begin{eqnarray}
K_1 &=& \frac{1.05}{\sqrt3 \pi^4 T_c^2 t} \ln \frac{t}{T_c} + \mathrm{subleading}\nonumber \\
K_2 &=& \frac{1.05}{ \sqrt{3} \pi^4 T_c^2 t} \ln \frac{t}{T_c} + \mathrm{subleading} = K_1\nonumber\\
K_3 &=& \frac{1.05 \bigg(\cos (\frac{2\pi}{3}) + 2 \bigg)}{\sqrt3 \pi^4 t T_c^2} \ln \frac{t}{T_c}  + \mathrm{subleading} \nonumber\\
&=& K_1 \bigg(\cos(\frac{2\pi}{3}) + 2 \bigg).
\end{eqnarray}
Note that there are two processes
 which  contribute to the coefficient $K_3$. The processes are represented diagrammatically in Fig.(\ref{fig: diag}). The process shown in Fig.(\ref{fig: diag},a) is sensitive to the chirality of the superconducting order parameter, because of the dependence on the phase difference between different saddle points, and gives rise to the $\cos (\frac{2\pi}{3})$ term.
  Because $\cos (2\pi/3) = -1/2$, this process
  gives rise to an effective attraction between superconductivity and spin density waves. This effective attraction is,
   however,
   outweighed by a larger (chirality independent) repulsion between
    the two order parameters,
    coming from the processes shown in Fig.(\ref{fig: diag},b).
     The prefactors in our case are such that
     $K_3 = \frac32 K_1 >0$.
     Comparing $K_1K_2$ and $K^2_3$ we see that in the case of doped graphene $K_1K_2 - K_3^2 <0$, so that co-existence is disfavored.
    The system only allows one order parameter to exist, even when $T_c = T_N$. 
     A direct 
     second order transition between superconducting and SDW orders is Landau forbidden, since the symmetry group of one ordered phase is not a subgroup of the symmetry group of the other ordered phase, and as a result the transition separating the
    region when $\Delta \neq 0, M=0$ from the region where $\Delta =0, M \neq 0$ is expected to be first order. (
    Although
     we cannot exclude a non-Landau continuous transition between the two ordered states.)

The fact that 
$T_c \neq T_N$
 makes co-existence
 even
  less likely.
  We therefore conclude that
  {\it there is no co-existence of superconducting and
  SDW
   order in
  doped
  graphene}.

In pnictides, the structure of $K_1, K_2, K_3$ is quite similar~\cite{Schmalian} (modulo that there is no $\log t/T_c$ term), but the argument of $\cos$ in $K_3$ is the
 phase difference between the gaps on hole and electron FSs. For $s^{+-}$ superconductivity, the argument is $\pi$, in which case $K_3 = K_1 =K_2$.
 Then, $K_1 K_2 = K^2_3$, and one has to include subleading terms to verify whether the two orders can co-exist. 
  The subleading terms 
  are the ones which  break the nesting between hole and electron pockets, and the analysis shows~\cite{Vorontsov,Schmalian} that
 $s^{+-}$ superconducting and SDW orders co-exist in some range of parameters. In graphene, 
  the argument of $\cos$ is $2\pi/3$, and
  such co-existence does not occur.

To conclude, we have demonstrated that superconductivity and spin density wave order are mutually exclusive in graphene doped near the $M$ point of the Brillouin zone
(a Van Hove doping).
 Sufficiently close to the Van Hove point, we expect to see pure chiral superconductivity, and somewhat away from the Van Hove point we expect to see pure spin density wave order. The results stand in stark contrast to pnictides, where there can be co-existence between spin density waves and superconductivity.

We acknowledge useful conversations with C. Batista,  G-W. Chern, FaWang, R. Fernandes,
 D-H Lee, I. Martin, J. Schmalian, and R. Thomale.    A.V.C. is supported by
NSF-DMR-0906953.

\begin{widetext}
\section{Appendix}

\subsection{Evaluating $K_1$}
We start with the expression
\begin{equation}
K_1 = 3 T_c \sum_{\omega_n} \int \frac{d^2k}{(2\pi)^2}  \big(F^+(\omega_n, \vec{k})F^-(\omega_n, \vec{k})\big)^2
= 3 T_c \sum_{\omega_n} \int \frac{d^2k}{(2\pi)^2}  \frac{1}{\big(\omega_n^2 + \frac{9t^2}{16}(3 k_x^2 - k_y)^2 \big)^2}
\end{equation}
Where the sum goes over all Matsubara frequencies, and the integral goes over all wavevectors $\vec{k}$ up to a UV cutoff of $|\vec{k}| \sim O(1)$, at which point the dispersion relation changes. The UV cutoff must be retained in the integrals, because the integrals are log divergent in the UV if we ignore the cutoff   (the integrals are convergent in the infrared at nonzero temperature). We now scale out $\omega_n$, and define the rescaled co-ordinates $x = \sqrt{3t/(4\omega_n)} k_x$, $y = \sqrt{3t/(4\omega_n)} k_y$. The expression for $K_1$ can then be recast as
\begin{equation}
K_1 = 3 T_c \sum_{\omega_n} \frac{1}{3 \pi^2 t |\omega_n|^3}\int_{-\sqrt{t/T_c}}^{\sqrt{t/T_c}} \frac{dx dy}{\big(1+(3x^2 - y^2)^2 \big)^2}
\end{equation}
We now integrate first over $-\infty < y < \infty$, and then over $-\sqrt{t/T_c}<x<\sqrt{t/T_c}$, and expand the resulting expression to leading order in large $t/T_c$ (the manipulations are all done on Mathematica). We obtain the result
\begin{equation}
K_1 =  T_c \sum_{\omega_n} \frac{1}{ \pi^2 t |\omega_n|^3} \bigg(\frac{\pi}{2 \sqrt{3}} \ln \frac{t}{T_c} + {\rm subleading}\bigg).
\end{equation}
The same result is obtained, with logarithmic accuracy, if we first integrate over $-\infty < x < \infty$, and then over $-\sqrt{t/T_c}<y<\sqrt{t/T_c}$. We now recall that $\omega_n = (2n+1) \pi T_c$, and that $\sum_n \frac{1}{|n+1/2|^3} = 14 \zeta(3) \approx 16.8$ (the sum may again be evaluated on Mathematica). Thus we obtain, with logarithmic accuracy, the result quoted in the main text, i.e.
\begin{equation}
K_1 =  \frac{14 \zeta(3)}{16 \sqrt{3} \pi^4  t T_c^2}\ln \frac{t}{T_c} \approx  \frac{1.05}{\sqrt{3} \pi^4  t T_c^2}\ln \frac{t}{T_c}
\end{equation}

\subsection{Evaluating $K_2$}
The co-efficient $K_2$ was evaluated already in \cite{SDW}. We can write $K_2 = 6 Z_1 + 12 Z_2$, where $Z_1$ and $Z_2$ are coefficients that were defined in \cite{SDW} and calculated in the supplement to \cite{SDW}. (Note that there is an overall factor of 2 relative to \cite{SDW}, which comes about because we have doubled the number of degrees of freedom in going to the Nambu spinor representation. However, this overall factor of 2 multiplies all terms in our free energy, and thus has no physical significance). In \cite{SDW}, it was shown that the term $Z_1$ was larger than $Z_2$ by a factor of $\ln t/T_c$, which is a large number at weak coupling. Thus, we can neglect $Z_2$ with logarithmic accuracy, and say $K_2 = 6 Z_1$.  The co-efficient $Z_1$ was calculated in \cite{SDW}, however, the calculation there had a factor of 2 error, which was unimportant for the physics considered in \cite{SDW} but is important here. Therefore, we redo the calculation of $Z_1$.

We wish to evaluate
\begin{equation}
Z_1 = T \sum_{\omega_n} \int \frac{d^2k}{(2\pi)^2} F^+(\vec{k}, \omega_n)^2 F^+(\vec{k+Q}, \omega_n)^2
\end{equation}
The integral over the Brillouin zone is dominated by those values of $\vec{k}$ where both Green functions correspond to states near a saddle point. Expanding the energy about the saddle points, we rewrite the integral as
\begin{equation}
Z_1 \approx T \sum_{\omega_n} \int \frac{d^2k}{(2\pi)^2} \frac{1}{\big(i\omega_n - \frac{3t_1}{4}(3k_x^2 - k_y^2)\big)^2\big(i\omega_n - \frac{3t_1}{4} 2k_y(k_y - \sqrt{3}k_x)\big)^2}
\end{equation}
Where the integral is understood to have a UV cutoff for $\vec{k}$ of order $1$.
We now define $a = \sqrt{3t_1/4} (k_y - \sqrt{3} k_x)$ and $b = \sqrt{3t_1/4}(k_y + \sqrt{3} k_x)$, and rewrite the above integral as
\begin{equation}
Z_1 = T \sum_{\omega_n} \frac{2}{3\sqrt{3} t_1}\int^{\sqrt{t_1}}_{-\sqrt{t_1}} \frac{da db}{(2\pi)^2} \frac{1}{\big(i\omega_n+ ab\big)^2\big(i\omega_n - a (a+b)\big)^2}
\end{equation}
We now define $x = ab$ and rewrite the integral as
\begin{equation}
Z_1 = T \sum_{\omega_n} \frac{2}{3\sqrt{3} t_1}\int^{\sqrt{t_1}}_{-\sqrt{t_1}} \frac{da}{2\pi} \frac{1}{|a|} \int_{-\sqrt{t_1} a}^{\sqrt{t_1}a} \frac{dx}{2\pi} \frac{1}{\big(i\omega_n + x\big)^2\big(i\omega_n - a^2 - x\big)^2}
\end{equation}
We now assume $T_N \ll t_1$ (which should certainly be the case for
 weak/moderate
 coupling). In this limit, we can perform the integral over $x$ approximately, using the Cauchy integral formula, to get
\begin{equation}
Z_1 = T \sum_{\omega_n} \frac{2}{3\sqrt{3} t_1}\int^{\sqrt{t_1}}_{-\sqrt{t_1}} \frac{da}{2\pi} \frac{1}{|a|} \frac{2 i \sign{\omega_n}}{(a^2 - 2 i \omega_n)^3} =  T \sum_{\omega_n} \frac{4}{3\sqrt{3} t_1}\int^{\sqrt{t_1}}_{-\sqrt{t_1}} \frac{da}{2\pi} \frac{1}{|a|} \frac{ i \sign{\omega_n (a^2 + 2 i \omega_n)^3}}{(a^4 + 4 \omega_n^2)^3}
\end{equation}
The imaginary part of the above integral is odd in $\omega$ and hence vanishes upon performing the Matsubara sum to leave an integral that is purely real
\begin{equation}
Z_1 = T \sum_{\omega_n} \frac{8 |\omega_n|}{3\sqrt{3} t_1}\int^{\sqrt{t_1}}_{-\sqrt{t_1}} \frac{da}{2\pi} \frac{1}{|a|} \frac{4 \omega_n^2 - 3 a^4}{(a^4 + 4 \omega_n^2)^3} \approx T \sum_{\omega_n} \frac{8 |\omega_n|}{3\sqrt{3} t_1}\int^{\sqrt{t_1}}_{-\sqrt{t_1}} \frac{da}{2\pi} \frac{1}{|a|} \frac{4 \omega_n^2}{(a^4 + 4 \omega_n^2)^3}
 \end{equation}
 with logarithmic accuracy. Performing the integral over $a$ (again with logarithmic accuracy) gives
 \begin{equation}
 Z_1 \approx T \sum_{\omega_n} \frac{1}{12 \pi \sqrt{3} t_1} \frac{1}{|\omega_n|^3} \ln \frac{t_1}{\omega_n} = \frac{1}{96 \pi^4 \sqrt{3} T_N^2 t_1} \big(16.8 \ln \frac{t_1}{2\pi T}  + 10.5 \big) \approx  \frac{16.8 \ln \frac{t_1}{T_N}}{96 \pi^4 \sqrt{3} T_N^2 t_1}
 \end{equation}
 Where we take $\omega_n = 2\pi (n+1/2) T_N$, $T = T_N$ and perform the discrete sum on mathematica. The error in the supplement to \cite{SDW} was in the last line of the calculation.

\subsection{Evaluating $K_3$}
There are two distinct contributions to $K_3$, and we evaluate both in turn. We can write $K_3 = K_3^a + K_3 ^b$, where
\begin{eqnarray}
K_3^a &= 6 T \sum_{\omega_n} \int \frac{d^2k}{(2\pi)^2} & F^+(\omega_n, \vec{k})F^-(\omega_n, \vec{k})F^+(\omega_n, \vec{k+Q_1})F^-(\omega_n, \vec{k+Q_1}) \cos(\theta_{\vec{k}} - \theta_{\vec{k+Q}}) \nonumber\\
K_3^b&= 6 T \sum_{\omega_n} \int \frac{d^2k}{(2\pi)^2} &  F^+(\omega_n, \vec{k})^2F^-(\omega_n, \vec{k})F^+(\omega_n, \vec{k+Q_1}) + (F^+ \leftrightarrow F^-)
\end{eqnarray}
The first contribution $K_3^a$ comes from processes of the form shown in Fig.(\ref{fig: diag},a), and is sensitive to the chirality of the superconducting order parameter (it depends on the difference in the phase of the superconducting order parameter at different points on the Fermi surface). This process leads to an attraction between chiral superconductivity and spin density waves The second contribution $K_3^b$ comes from processes of the form shown in Fig.(\ref{fig: diag},b), and is insensitive to the chirality of the superconducting order parameter. This process leads to a repulsion between any kind of superconductivity and spin density waves. The second process dominates (because of purely numerical prefactors), so superconductivity and spin density waves do repel - but the repulsion is too weak to prevent co-existence.

Let us first calculate $K_3^a$. For $d+id$ pairing, we have $\theta_{\vec{k}} - \theta_{\vec{k+Q}} = 4\pi/3$. Thus, we have
\begin{eqnarray}
K_3^a &= 6 T_c \sum_{\omega_n} \int \frac{d^2k}{(2\pi)^2} & F^+(\omega_n, \vec{k})F^-(\omega_n, \vec{k})F^+(\omega_n, \vec{k+Q_1})F^-(\omega_n, \vec{k+Q_1}) \cos\frac{4\pi}{3} \nonumber\\
&= 6 T_c \cos \frac{4\pi}{3} \sum_{\omega_n} \int \frac{d^2k}{(2\pi)^2} & \frac{1}{\big(\omega_n^2 + \frac{9t^2}{16}(3k_x^2-k_y^2)^2\big)\big(\omega_n^2 +  \frac{9t^2}{16} 4 k_y^2 (k_y-\sqrt3 k_x)^2 \big)}.
\end{eqnarray}
We scale out $\omega_n$ and define rescaled variables $x = \sqrt{3t/(4\omega_n)} k_x$, $y = \sqrt{3t/(4\omega_n)} k_y$. The expression for $K^a_3$ can then be recast as
\begin{equation}
K_3^a =\cos\bigg( \frac{4\pi}{3}\bigg) \sum_{\omega_n} \frac{ 2 T_c }{\pi^2 t |\omega_n|^3} \int_{-\sqrt{t/T_c}}^{\sqrt{t/T_c}}\frac{dx dy}{\big(1 + (3x^2-y^2)^2\big)\big(1 +  4 y^2 (y-\sqrt3 x)^2 \big)}.
\end{equation}
We define the new coordinates $a = y - \sqrt3 x$, $b = y + \sqrt3 x$, and hence re-express the above integral (with logarithmic accuracy) as
\begin{equation}
K_3^a =\cos\bigg( \frac{4\pi}{3}\bigg) \sum_{\omega_n}\frac{ T_c }{\sqrt3 \pi^2 t |\omega_n|^3} \int_{-\sqrt{t/T_c}}^{\sqrt{t/T_c}}  \frac{da db}{\big(1 + a^2 b^2\big)\big(1 +  a^2 (a+b)^2 \big)}.
\end{equation}
We integrate over $-\sqrt{t/T_c}< b < \sqrt{t/T_c}$ (on Mathematica), and expand the resulting expression to leading order in large $t/T_c$. This leads to the expression
\begin{equation}
K_3^a =\cos\bigg( \frac{4\pi}{3}\bigg) \sum_{\omega_n} \frac{ T_c }{\sqrt3 \pi^2 t |\omega_n|^3} \bigg(\int da \frac{\pi}{2|a|(1+a^4/4)} \bigg)
\end{equation}
It should be remembered that the expansion in large $t/T_c$ is valid only for $a^2 t/T_c \gg 1$, thus the above integral implicitly carries an infrared cutoff on the scale $a \approx \sqrt{t/T_c}$. Performing the integral with this infrared cutoff, we obtain the expression
\begin{equation}
K_3^a =\cos\bigg( \frac{4\pi}{3}\bigg) \sum_{\omega_n} \frac{ T_c }{2 \sqrt3 \pi t |\omega_n|^3} \ln \frac{t}{T_c}
\end{equation}
Performing the summation over $n$ on Mathematica, as before, we obtain
\begin{equation}
K_3^a =\cos\bigg( \frac{4\pi}{3}\bigg) \frac{14 \zeta{3}}{16 \sqrt3 \pi^4 t T_c^2} \ln \frac{t}{T_c} \approx \frac{-1}{2} \frac{16.8}{16 \sqrt3 \pi^4 t T_c^2} \ln \frac{t}{T_c} =  \frac{-1.05}{2 \sqrt3 \pi^4 t T_c^2} \ln \frac{t}{T_c} = - \frac12 K_1
\end{equation}
Note the crucial minus sign that comes from the chirality sensitive $\cos$ factor - this particular term represents an attraction between magnetism and chiral superconductivity.

We now turn our attention to the second term, $K_3^b$. We have
\begin{eqnarray}
K_3^b&= 6 T_c \sum_{\omega_n} \int \frac{d^2k}{(2\pi)^2} &  F^+(\omega_n, \vec{k})^2F^-(\omega_n, \vec{k})F^+(\omega_n, \vec{k+Q_1}) + (F^+ \leftrightarrow F^-) \nonumber\\
&= 6 T_c \sum_{\omega_n} \int \frac{d^2k}{(2\pi)^2} &  \frac{-\bigg(\big(i\omega_n + \frac{3t}{4} (3k_x^2-k_y^2)\big)\big(i\omega_n + \frac{3t}{4} 2k_y (k_y-\sqrt3 k_x)\big)+ \big(\omega_n \rightarrow - \omega_n\big)\bigg)}{\big(\omega_n^2 + \frac{9t^2}{16}(3k_x^2-k_y^2)^2\big)^2\big(\omega_n^2 +  \frac{9t^2}{16} 4 k_y^2 (k_y-\sqrt3 k_x)^2 \big)}\nonumber\\
&= 12 T_c \sum_{\omega_n} \int \frac{d^2k}{(2\pi)^2} &  \frac{\omega_n^2 - \frac{9t^2}{16}(3k_x^2-k_y^2)2k_y(k_y - \sqrt{3} k_x)}{\big(\omega_n^2 + \frac{9t^2}{16}(3k_x^2-k_y^2)^2\big)^2\big(\omega_n^2 +  \frac{9t^2}{16} 4 k_y^2 (k_y-\sqrt3 k_x)^2 \big)}
\end{eqnarray}
 Again,
  we scale out $\omega_n$
   and define the rescaled variables $x = \sqrt{3t/(4\omega_n)} k_x$, $y = \sqrt{3t/(4\omega_n)} k_y$. The expression for $K^b_3$ can then be recast as
\begin{equation}
K_3^b =  \sum_{\omega_n}  \frac{4T_c }{\pi^2 t |\omega_n|^3} \int_{-\sqrt{t/T_c}}^{\sqrt{t/T_c}} dx dy \frac{1 - (3x^2-y^2)2y(y-\sqrt3 x)}{\big(1 + (3x^2-y^2)^2\big)^2\big(1 +  4 y^2 (y-\sqrt3 x)^2 \big)}
\end{equation}
We define the new coordinates $a = y-\sqrt3 x$, $b = y + \sqrt3 x$, and hence re-express the above integral (with logarithmic accuracy) as
\begin{equation}
K_3^b =  \sum_{\omega_n}  \frac{2T_c }{\sqrt3 \pi^2 t |\omega_n|^3} \int_{-\sqrt{t/T_c}}^{\sqrt{t/T_c}} da db \frac{1 + a^2 b (a+b)}{\big(1 + a^2 b^2\big)^2\big(1 +  a^2 (a+b)^2 \big)}
\end{equation}
We integrate over $-\sqrt{t/T_c}< b < \sqrt{t/T_c}$ (on Mathematica), and expand the resulting expression to leading order in large $t/T_c$. This leads to the expression
\begin{equation}
K_3^b =  \sum_{\omega_n}  \frac{2T_c }{\sqrt3 \pi^2 t |\omega_n|^3} \bigg(\int da \frac{\pi}{2 |a| (1+a^4/4)^2} \bigg)
\end{equation}
It should be remembered that the expansion in large $t/T_c$ is valid only for $a^2 t/T_c \gg 1$, thus the above integral implicitly carries an infrared cutoff on the scale $a \approx \sqrt{t/T_c}$. Performing the integral with this infrared cutoff, we obtain the expression
\begin{equation}
K_3^b =\sum_{\omega_n} \frac{ T_c }{\sqrt3 \pi t |\omega_n|^3} \ln \frac{t}{T_c} = \frac{2.01}{\sqrt3 \pi^4 t T_c^2} \ln \frac{t}{T_c} = 2 K_1
\end{equation}
Putting things together, we have
\begin{equation}
K_3 = K_3^a+K_3^b = K_1 \big(\cos \frac{2\pi}{3} + 2) = K_1 (-\frac12 +2) = \frac32 K_1
\end{equation}
 quoted in the main text.
\end{widetext}

\end{document}